\documentclass[twoside]{article}
\usepackage{Proc_NTSE_14}
\usepackage{amssymb,amsmath}

\newcommand{\bra}[1]{\langle {#1} |}
\newcommand{\ket}[1]{| {#1} \rangle}
\newcommand{\inproduct}[2]{\langle #1 | #2 \rangle}

\begin{document}
\thispagestyle{plain}
%Please use the command \publref{myfilename} to print the reference to your proceedings contribution 
%at the bottom of the page where myfilename should be replaced by the name of your LaTeX file
%(e.g., use the command \publref{Johns} if the LaTeX file of your contribution called Johns.tex):
\publref{Nakatsukasa}

\begin{center}
{\Large \bf \strut
%Insert the title of your contribution here
Time-dependent density-functional calculation of nuclear response functions
% Template for  contributions to the proceedings of the NTSE-2014 conference
\strut}\\
\vspace{10mm}
{\large \bf 
%Insert the authors here. Use upper indexes a, b, c, etc., to bind authors with their addresses
% as shown below.
Takashi Nakatsukasa}
%$^{a,b,c}$, I. I. Author$^{c,d}$ and  Third C. Author{$^d$}}
\end{center}

\noindent{% Insert the addresses  here.
%\small $^a$\it The first address} \\
%{\small $^b$\it The second address } \\
%{\small $^c$\it The third address} \\
% {\small $^d$\it The fourth address}
\begin{center}
 {\small Center for Computational Sciences, University of Tsukuba,
Tsukuba 305-8577, Japan\\
 and \\
RIKEN Nishina Center, Wako 351-0198, Japan}
\end{center}

%The next command defines running titles:
\markboth{
%Put here the list of authors that will be displayed in running titles:
Takashi Nakatsukasa}
%First Author, I. I. Author and Third C. Author}
{%Put here the short title of your contribution that will be displayed in running titles:
TDDFT calculation of nuclear response functions} 

\begin{abstract}
Basic issues of the time-dependent density-functional theory
are discussed, especially on the real-time calculation of
the linear response functions.
Some remarks on the derivation of the time-dependent Kohn-Sham
equations and on the numerical methods are given.
\\[\baselineskip] 
\end{abstract}

\section{Introduction}

The time-dependent density functional theory (TDDFT) provides us with
a practical tool to study quantum dynamics of many-body systems.
It is conceptually very similar to the one known as the
time-dependent Hartree-Fock (TDHF) theory with
a density-dependent effective interaction in nuclear physics \cite{Neg82}.
Although it is much more feasible than directly treating many-body
wave functions of many-particle systems,
the studies of full dynamics taking into account
both the mean-field and the pairing correlations are still computationally
highly challenging, even at present.
In this paper, for simplicity, I will concentrate the discussion on
the time-dependent Kohn-Sham (TDKS) equations,
without the Bogoliubov-type extension including pair densities.

Recently, there are a number of developments in the studies of nuclei
with the density functional approaches.
For these, there are 
recent review papers \cite{BHR03,Nak12,Sim12,MRSU14}.
Thus, I do not intend here to review all these developments.
Instead, I would like to present some issues which are not well addressed in
published articles.
The first issue, presented in Sec. \ref{sec: derivation},
is the derivation of the time-dependent
Kohn-Sham equations
based on the time-dependent variational principle.
Exactly the same argument is applicable to the variational derivation of
the time-dependent Hartree-Fock equations.
Especially, I would like to clarify that the gauge degrees of freedom
naturally emerge in the proper derivation.
This may be trivial to some readers,
however, I think it is not so for non-practitioners.
It may be also useful for students.
Then, in Sec. \ref{sec: numerical}, I will present some practical issues
on numerical calculations, such as
choice of the gauge functions, some speed-up techniques, etc.

\section{Remarks on the time-dependent Kohn-Sham equations}
\label{sec: derivation}

\subsection{Time-dependent variational principle}

It is well-known that the time-dependent Kohn-Sham equations
can be obtained using the time-dependent variational principle \cite{KK76}.
In literature, we often find the following arguments:
Starting from the action integral
\begin{eqnarray}
\label{action_1}
S &\equiv& \int_{t_i}^{t_f} \left\{
\bra{\Psi_D(t)} i\frac{\partial}{\partial t} \ket{\Psi_D(t)}
- E[\rho(t)] 
\right\} dt \\
\label{action_2}
&=&
\int_{t_i}^{t_f} \left\{
\sum_{i=1}^N \bra{\psi_i(t)} i\frac{\partial}{\partial t} \ket{\psi_i(t)}
- E[\rho(t)] 
\right\} dt ,
\end{eqnarray}
the stationary action principle, $\delta S=0$, leads to the
time-dependent Kohn-Sham (TDKS) equations
\begin{equation}
\label{TDKS_eq}
i\frac{\partial}{\partial t} \ket{\psi_i(t)} = h[\rho(t)] \ket{\psi_i(t)} ,
\quad i=1,\cdots,N .
\end{equation}
Here, $E[\rho]$ is the energy density functional and
$\ket{\Psi_D(t)}$ is the time-dependent Slater determinant,
\begin{equation}
\label{SD}
\ket{\Psi_D(t)} = \frac{1}{\sqrt{N!}}
\textrm{det}\{ \ket{\psi_i(t)}_j \}_{i,j=1,\cdots, N} .
\end{equation}
The Kohn-Sham (single-particle) Hamiltonian is formally defined by
\begin{equation}
h[\rho] \ket{\psi_i} = \frac{\delta E}{\delta \bra{\psi_i}} .
\end{equation}
In the case of TDHF with an effective Hamiltonian, the energy density
functional is given by the expectation value of the
Hamiltonian as $E[\rho(t)] = \bra{\Psi_D(t)} H \ket{\Psi_D(t)}$.
In general, $E[\rho]$ can be a more general functional of one-body density
$\rho$ in the TDDFT.

Since the TDKS equations (\ref{TDKS_eq}) are so common in literature,
I think many people take them for granted.
However, it is somewhat strange that
we have reached the equations which can uniquely
determine the Kohn-Sham orbitals, because
the Slater determinant $\ket{\Psi_D(t)}$ is invariant under the
unitary transformation among the occupied orbitals.
Namely,  the same Slater determinant $\ket{\Psi_D(t)}$ can be expressed by
different orbitals,
$\ket{\psi'_i(t)} = \sum_{j=1}^N U_{ij}(t) \ket{\psi_j(t)}$,
as Eq. (\ref{SD}), where $\{ U_{ij}(t) \}$ is
an arbitrary time-dependent unitary matrix.
Thus, the Kohn-Sham orbitals have gauge degrees of freedom associated
with the $U(N)$ transformation.

Apparently, the TDKS equation (\ref{TDKS_eq}) uniquely determines
the time evolution of each single-particle orbital $\ket{\psi_i(t)}$.
Since we have not imposed any gauge fixing
condition when we derived Eq. (\ref{TDKS_eq})
from the stationary action principle, $\delta S=0$,
for Eq. (\ref{action_2}),
the Kohn-Sham (single-particle) orbitals should not be unique.

In fact, to my opinion,
the derivation above is not satisfactory because we have used the
orthonormal condition among the orbitals,
$\inproduct{\psi_i(t)}{\psi_j(t)}=\delta_{ij}$,
to obtain Eq. (\ref{action_2}) from Eq. (\ref{action_1}).
Therefore, the full variation with respect to
each $\bra{\psi_i(t)}$ should not be taken.
I think that the proper derivation is either
(i) the orthonormal relations are not assumed in the first place,
or (ii) the variational space is restricted by the constraints
$\inproduct{\psi_i(t)}{\psi_j(t)}=\delta_{ij}$.
In the following,
I would like to present these proper derivations of the TDKS equations
and show that the gauge degrees of freedom naturally appear.

\subsection{Derivation of TDKS equations (1)}

To allow us to take full variation with respect to $\psi_i(t)$,
we should not assume the orthonormal relation among $\{ \ket{\psi_i} \}$.
Let us derive the equations,
starting from the action 
\begin{equation}
\label{action_3}
S \equiv \int_{t_i}^{t_f} \left\{
\frac{
\bra{\Psi_D(t)} i\frac{\partial}{\partial t} \ket{\Psi_D(t)}}
{\inproduct{\Psi_D(t)}{\Psi_D(t)}}
- E[\rho(t)] 
\right\} dt .
\end{equation}

In order to perform calculation of the functional derivatives,
some formulae,
which are well-known in the generator coordinate method (GCM)
\cite{RS80}, are very helpful.
First, it is useful to define the overlap matrix,
\begin{equation}
B_{ij}(t)\equiv \inproduct{\psi_i(t)}{\psi_j(t)} ,
\quad i,j=1,\cdots, N ,
\end{equation}
which leads to the following expressions for the norm and the
time derivative.
\begin{eqnarray}
{\inproduct{\Psi_D(t)}{\Psi_D(t)}}
&=& \textrm{det}B , \\
\bra{\Psi_D(t)} i\frac{\partial}{\partial t} \ket{\Psi_D(t)}
&=& \textrm{det}B \sum_{ij} \bra{\psi_i(t)} 
i\frac{\partial}{\partial t} \ket{\psi_j(t)}
\left(B^{-1}\right)_{ji} .
\end{eqnarray}
Hereafter, the summation $\sum_i$ means the summation with respect
to the occupied (hole) orbitals, $i=1,\cdots, N$,
and the time-dependent overlap matrix $B(t)$ is simply denoted as $B$,
for simplicity.
Using the cofactor expansion of the inverse matrix $B^{-1}$,
we can prove that
\begin{equation}
\frac{\delta \left( B^{-1}\right)_{ij}(t')}{\delta \bra{\psi_k(t)}}
= - \left( B^{-1}\right)_{ik} \sum_l
 \ket{\psi_l} \left( B^{-1}\right)_{lj}
 \delta(t-t') .
\end{equation}
In the same manner, the one-body density matrix can be written as
\begin{equation}
\label{rho}
\rho(t) = \sum_{ij} \ket{\psi_i(t)}
\left( B^{-1}\right)_{ij}
\bra{\psi_j(t)} .
\end{equation}
Then, the derivative of $E[\rho]$ with respect to the bra state
$\bra{\psi_k(t)}$ becomes
\begin{equation}
\frac{\delta E[\rho]}{\delta \bra{\psi_k(t)}}
= h[\rho(t)] \sum_j \ket{\psi_j(t)} \left( B^{-1}\right)_{jk} .
\end{equation}
Now, it is easy to derive the TDKS equations
\begin{equation}
\label{TDKS_eq_2}
\left(
1-\sum_l \ket{\psi_l(t)}\left(B^{-1}\right)_{lj}\bra{\psi_j(t)}
\right)
\left( i\frac{\partial}{\partial t} - h[\rho(t)] \right)
\sum_k \ket{\psi_k(t)} \left(B^{-1}\right)_{ki}
= 0 .
\end{equation}
This looks different from the well-known form of Eq. (\ref{TDKS_eq}).

We may simplify Eq. (\ref{TDKS_eq_2}) by assuming that
the orbitals are orthonormal at a certain time $t$,
$\inproduct{\psi_i(t)}{\psi_j(t)}=\delta_{ij}$.
In this case, we have $B_{ij}(t)=(B^{-1})_{ij}=\delta_{ij}$.
Then, Eq. (\ref{TDKS_eq_2}) can be written as
\begin{equation}
\label{TDKS_eq_3}
\left(
1-\sum_j \ket{\psi_j(t)}\bra{\psi_j(t)}
\right)
\left( i\frac{\partial}{\partial t} - h[\rho(t)] \right)
\ket{\psi_i(t)} = 0 ,
\end{equation}
at time $t$.
This means that the states
\begin{equation}
\left( i\frac{\partial}{\partial t} - h[\rho(t)] \right) \ket{\psi_i(t)}
\end{equation}
do not contain the particle (unoccupied) orbitals at time $t$.
In other words,
they can be expanded in terms of the hole (occupied) orbitals only.
\begin{equation}
\label{TDKS_eq_4}
i\frac{\partial}{\partial t} \ket{\psi_i(t)} 
= h[\rho(t)] \ket{\psi_i(t)}
+ \sum_k \lambda_{ij}(t) \ket{\psi_j(t)} .
\end{equation}
Although $\lambda_{ij}(t)$ are in principle arbitrary,
choosing the Hermitian matrix $\lambda_{ik}(t)$
will conserve the orthonormal relation among
the orbitals $\inproduct{\psi_i(t+\Delta t)}{\psi_j(t+\Delta t)}=\delta_{ij}$.
Therefore, provided that $\lambda_{ij}(t)$ being Hermitian,
Eq. (\ref{TDKS_eq_4}) can be true for any time $t$.
They can be regarded as a general form of
the TDKS equations.
Here, the time-dependent Hermitian matrix $\lambda_{ij}(t)$ is
a kind of gauge function for fixing the Kohn-Sham orbitals.

Equation (\ref{TDKS_eq_4}) is also consistent with
the well-known form of the equation for the one-body density matrix.
Since the orthonormal relation is kept all the time,
the density matrix of Eq. (\ref{rho}) can be simplified by
assuming $B^{-1}_{ij}=\delta_{ij}$.
Then, the time derivative of $\rho(t)$ can be calculated as
\begin{eqnarray}
i\frac{\partial \rho}{\partial t} &=& \sum_i
\left( h \ket{\psi_i} + \sum_j \lambda_{ij} \ket{\psi_j} \right) \bra{\psi_i}
-\sum_i
\ket{\psi_i}\left( \bra{\psi_i} h + \sum_j \bra{\psi_j} \lambda_{ij}^* \right)
\\
&=& \sum_i \left( h\ket{\psi_i}\bra{\psi_i}
- \ket{\psi_i}\bra{\psi_i} h \right) \\
&=& \left[ h[\rho(t)],\rho(t) \right] .
\end{eqnarray}

\subsection{Derivation of TDKS equations (2)}

We saw in the previous section that the calculation of the functional
derivative of the action $S$ in Eq. (\ref{action_3}) is rather
tedious.
The use of Lagrange multipliers may greatly facilitate this calculation.
One of the great advantages of the Lagrange multipliers is that,
when we impose the constraints in terms of the Lagrange multipliers,
we may simplify the functionals (functions) by using the constraints
{\it before variations}.
Now, we can use the action $S$ in the simple form of Eq. (\ref{action_2}) but
with the Lagrange multipliers to impose the constraints of the
orthonormal relation $\inproduct{\psi_i(t)}{\psi_j(t)} = \delta_{ij}$.
\begin{eqnarray}
\delta \left\{
S - \sum_{ij} \lambda_{ij}(t)
 \left( \inproduct{\psi_i(t)}{\psi_j(t)} - \delta_{ij} \right)
\right\}
= 0 .
\end{eqnarray}
The variation immediately leads to Eq. (\ref{TDKS_eq_4}).
The form of Eq. (\ref{TDKS_eq_4}) with the Hermitian matrix $\lambda(t)$
can be regarded as a general form of the TDKS equations.

Before ending this section, let us show that we can use
the constraint conditions to simplify the functions
before variation, when the Lagrange multipliers are utilized.
We consider here a problem to find the extrema of a function $F(\vec{x})$
with a constraint $g(\vec{x})=0$.
Using the Lagrange multiplier $\lambda$, it can be given by the
following variational form,
\begin{equation}
\label{delta_F}
\delta \left\{ F(\vec{x})-\lambda g(\vec{x}) \right\} = 0 
\quad\rightarrow\quad
{\nabla} F(\vec{x}) - \lambda {\nabla} g(\vec{x}) = 0 ,
\quad\textrm{with } g(\vec{x)}=0 .
\end{equation}
Namely, ${\nabla} F(\vec{x})$ is parallel to ${\nabla} g(\vec{x})$,
which is the condition of the extrema under the constraint of $g(\vec{x})=0$.
Now, let us assume that the functional form of $F(\vec{x})$ can be
modified (simplified) into $\tilde{F}(\vec{x})$
if we use the constraint $g(\vec{x})=0$.
\begin{equation}
\tilde{F}(\vec{x})={f}(\vec{x};g=0) ,
\end{equation}
where ${f}(\vec{x};g(\vec{x}))$ is a function of $\vec{x}$ and
$g(\vec{x})$, satisfying
${f}(\vec{x};g(\vec{x}))=F(\vec{x})$.
From these, we can rewrite Eq. (\ref{delta_F}) as
\begin{equation}
%\vec{\nabla} F(\vec{x})= 
{\nabla} \tilde{F}(\vec{x}) - 
\left(
\lambda -\frac{\partial {f}}{\partial g}
\right)
{\nabla} g(\vec{x}) 
= 0 ,
\quad\textrm{with } g(\vec{x})=0 .
\end{equation}
This means that ${\nabla} \tilde{F}(\vec{x})$
is also parallel to ${\nabla} g(\vec{x})$,
at the extrema with $g(\vec{x})=0$.
Therefore, it is identical to the following variation to find the extrema.
\begin{equation}
\delta \left\{ \tilde{F}(\vec{x})-\lambda g(\vec{x}) \right\} = 0 
\quad\rightarrow\quad
{\nabla} \tilde{F}(\vec{x}) - \lambda {\nabla} g(\vec{x}) = 0 ,
\quad\textrm{with } g(\vec{x})=0 .
\end{equation}
Thus, we can replace $F(\vec{x})$ by $\tilde{F}(\vec{x})$ for
the variational calculation with the Lagrange multiplier $\lambda$.
$\tilde{F}(\vec{x})=F(\vec{x})$ where $g(\vec{x})=0$
is satisfied.
An extension of the present argument to the case of multiple constraints
is straightforward.

\section{Remarks on the numerical calculations}
\label{sec: numerical}

The Kohn-Sham orbitals are evolved in time according to Eq. (\ref{TDKS_eq_4}).
As is shown in the previous section, there are gauge degrees of freedom
($\lambda_{ij}(t)$) we can choose arbitrarily.
Although the choice of the gauge should not affect the physical quantities,
the feasibility of numerical simulations sometimes depends on it.

\subsection{Preparation of the initial state}

In most applications, the initial state of the time evolution is
prepared by solving the static Kohn-Sham equations:
\begin{equation}
\label{KS_eq}
h[\rho] \ket{\psi_i} = \epsilon_i \ket{\psi_i} ,\quad
\textrm{ and } \quad
\rho=\sum_i \ket{\psi_i}\bra{\psi_i} .
\end{equation}
Of course, this is not the only way of constructing the ground-state
Kohn-Sham orbitals.
Again, the $U(N)$ gauge degrees of freedom exist for the ground state.
Nevertheless, they are somewhat special in the sense that
both the Hamiltonian $h[\rho]$ and the density $\rho$ are
diagonal in these orbitals.
They are often called ``canonical orbitals''.

To reach the ground state, the imaginary-time method is one of the
most prevalent methods in nuclear physics \cite{DFKW80}.
We start from given initial wave functions for $\ket{\psi_i^{(0)}}$
which are orthonormalized to each other.
Then, at the (n+1)-th iteration, 
the imaginary-time evolution of a small time step $\Delta t$
is calculated as
\begin{equation}
\ket{\psi_i^{(n+1)}}= \exp(-\Delta t \ h[\rho^{(n)}]) \ket{\psi_i^{(n)}}
\approx \left( 1 - \Delta t \ h[\rho^{(n)}] \right) \ket{\psi_i^{(n)}} ,
\end{equation}
where the Kohn-Sham Hamiltonian is constructed at the density of
$\rho^{(n)}$ which is defined by
\begin{equation}
\rho^{(n)} = \sum_i \ket{\psi_i^{(n)}} \bra{\psi_i^{(n)}} .
\end{equation}
At each iteration, the Gram-Schmidt orthonormalization must be performed.
This procedure converges to the solutions of Eq. (\ref{KS_eq}) from
the eigenstate of the lowest energy $\epsilon_1$ to
that of the $N$-th eigenvalue $\epsilon_N$.
You may also calculate the particle (unoccupied) states ($i>N$)
if you want.

An advantage of the imaginary-time method is that
it is a very stable iteration procedure to reach the convergence,
though it may require a large number of iterations.
Diagonalizing $h[\rho^{(n)}]$ in the space spanned by the set of states
$\{ \ket{\psi_i^{(n)}} \}_{i=1,\cdots, N}$ may speed up the convergence.
Sometimes, an additional damping factor associated with the kinetic
energy terms, $1/p^2$, could help to lower the energy quickly,
especially in the beginning stage of the iterations.

\subsection{Strength functions in the linear response}

The linear response in real time can be numerically realized, if we
slightly distort the ground-state density and start the time evolution.
The distortion is made by a weak external field, $V_\textrm{ext}(t)$.
The time profile of the external field determines the frequency range
contained in $V_\textrm{ext}(t)$.
\begin{equation}
V_\textrm{ext}(t)=\frac{1}{2\pi}\int \tilde{V}_\textrm{ext}(\omega)
 e^{-i\omega t} d\omega .
\end{equation}
One of the popular choices is the instantaneous field,
$V_\textrm{ext}(t)\propto \delta(t)$,
which correspond to the constant field in the frequency domain,
$\tilde{V}_\textrm{ext}(\omega)\sim V_0$.
An advantage of this instantaneous external field is that the
calculation of the single time evolution 
provides information on the all the frequency (energy) range.

The strength functions can be calculated in the real-time method
as follows.
Suppose that we want to calculate the strength function associated with
the one-body Hermitian operator $F$ for a system whose energy eigenstates
are denoted by $\ket{\Phi_n}$.
The initial state is constructed by applying the instantaneous
external field, $V_\textrm{ext}(t)=-\eta F \delta(t)$, at $t=0$,
which leads to $\ket{\Psi(t=0+)}=e^{i\eta{F}}\ket{\Psi_0}$.
Here, we adopt a small parameter of $\eta$ to numerically perform
the linear approximation.
The time-dependent state $\ket{\Psi(t)}$
can be decomposed in terms of $\ket{\Phi_n}$ as
\begin{equation}
\ket{\Psi(t)}=e^{-iHt}e^{i\eta F}\ket{\Psi_0}
= e^{-iE_0 t}\ket{\Phi_0} 
+ i\eta \sum_n e^{-iE_n t}\ket{\Phi_n}\bra{\Phi_n}F\ket{\Phi_0} 
+ {\cal O}(\eta^2) .
\end{equation}
Therefore, the calculation of the expectation value of $F$ leads to
\begin{equation}
\bra{\Psi(t)} F\ket{\Psi(t)} = \bra{\Phi_0} F \ket{\Phi_0}
+ 2\eta \sum_n |\bra{\Phi_n} F \ket{\Phi_0}|^2 \sin\{(E_n-E_0)t\}  .
\end{equation}
Then, the strength function is obtained by the Fourier transform.
\begin{eqnarray}
S(E;F) &\equiv& \sum_n 
|\bra{\Phi_n} F \ket{\Phi_0}|^2 \delta(E-(E_n-E_0)) \\
&=&
\frac{1}{\pi \eta} \int_0^\infty \sin(Et)
\{ \bra{\Psi(t)} F\ket{\Psi(t)} - \bra{\Phi_0} F \ket{\Phi_0} \} .
\label{strength_function}
\end{eqnarray}
In practice, it is impossible to perform the time evolution up to
$t=\infty$.
Usually we introduce an artificial damping (smearing) factor $\gamma$
to multiply the integrand of Eq. (\ref{strength_function}) by
$e^{-\gamma t/2}$, and stop integration at $t=T$.
The magnitude of the damping factor $\gamma$ is related to the
time duration $T$.
To obtain a smooth curve as a function of energy $E$,
we need to have $\gamma \gtrsim 2\pi/T$.

\subsection{Choice of the gauge functions}

The canonical orbitals of the ground state, defined by Eq. (\ref{KS_eq}),
should correspond to the stationary solutions of
the TDKS equations (\ref{TDKS_eq_4}).
However, apparently, off-diagonal parts of the gauge functions
$\lambda_{ij}(t)$ make
the solution not stationary, but the mixing among the hole orbitals
takes place in time.
When we choose the gauge,
$\lambda_{ij}(t)=-\epsilon_i\delta_{ij}$,
the static canonical orbitals of Eq. (\ref{KS_eq}) becomes stationary,
$\partial\psi_i/\partial t=0$.

In the real-time calculation of the linear response,
the state stays very close to the ground state.
Only a small part of the Kohn-Sham wave functions are fluctuating.
Therefore, it is convenient to adopt the
gauge same as the one above,
$\lambda_{ij}(t)=-\epsilon_i\delta_{ij}$.
Of course, the choice of the gauge is completely arbitrary and should
not affect the final results.
However, this choice has some numerical advantage, because
the time-dependent phase change of each Kohn-Sham orbital is minimized.

For calculation of nuclear dynamics beyond the linear regime, such as
simulation of heavy-ion collisions,
the choice of
$\lambda_{ij}(t)=-\epsilon_i\delta_{ij}$
is no longer advantageous.
Instead, we may adopt 
$\lambda_{ij}(t)=-\delta_{ij}\bra{\psi_i(t)} h[\rho(t)] \ket{\psi_i(t)}$,
for instance.

\subsection{Numerical applications}

In this article, we do not show results of numerical calculations.
I would like readers to refer to our previous papers
\cite{NY01,NY02-P2,NUY03-P,NY04-P1,NY04-P2,NY05,NY05-P1,NY07-P,NYI08-P,NY08-P,KYNNN09,Eba10,ENI12-P1,ENI12-P2,GN12-P,Nak12,Nak12-P,ENI14}.

\section{Acknowledgements}

This work is supported by Grant-in-Aid for Scientific Research
(Nos. 25287065 and 13327989).
Computational resources were partially provided by the HPCI Systems
Research Projects (hp120192).
\bibliographystyle{unsrt}
\bibliography{nuclear_physics,myself,current}

%\begin{thebibliography}{3}

%%Reference to a journal paper:
%\bibitem{Ref1} P. Maris, J. P. Vary and A. M. Shirokov, 
%Phys. Rev. C {\bf 79}, 014308 (2009).

%%Reference to a conference proceedings:
%\bibitem{Ref2} A. M. Shirokov, J. P. Vary and P. Maris, in {\em Proc.  27th
%Int. Workshop Nucl. Theory, Rila Mountains, Bulgaria, 23--28 June,
%2008}, edited by S. Dimitrova. Bulgarian Academy of Science, 2008, p. 205.

%% Reference to a book:
%\bibitem{Ref3} R. G. Newton, {\em Scattering theory of waves and
%particles,  2nd. ed.} Springer-Verlag, New York, 1982.

%\end{thebibliography}

\end{document}